\documentstyle[11pt,aaspp4]{article}

\def\arcm{\char'023\ ~}
\def\hub{\ifmmode H_\circ\else H$_\circ$\fi}
\def\kms{~km~s$^{-1}$\ }
\def\Hd{$\rm H\delta /\lambda 4045$}
\def\He2{$\rm H\delta /\lambda 4063$}
\def\Hg{$\rm H\gamma /\lambda 4325$}
\def\H8{$\rm H8/\lambda 3859$}
\def\HgHR{$\rm H\gamma _{HR}$}
\def\CaHR{$\rm Ca \ I_{HR}$}
\def\FeHR{$\rm 4045_{HR}$}
\journalid{337}{15 January 1989}
\articleid{11}{14}
\begin{document}

\title{Low-Luminosity Early-Type Galaxies in the Coma Cluster: Variations in
Spectral Properties }

\author{Nelson Caldwell\altaffilmark{1, 2}}
\affil{F.L. Whipple Observatory, Smithsonian Institution, Box 97, Amado AZ 85645}
\affil{Electronic mail: caldwell@flwo99.sao.arizona.edu}

\author{James A. Rose\altaffilmark{2}}
\affil{Department of Physics and Astronomy, University of North Carolina, Chapel Hill, NC 27599}
\affil{Electronic mail: jim@wrath.physics.unc.edu}


\altaffiltext{1}{Visiting Astronomer, Cerro Tololo Inter-American Observatory. 
CTIO is operated by AURA, Inc.\ under contract to the National Science
Foundation.} 
\altaffiltext{2}{Visiting Astronomer, Kitt Peak National Observatory.  KPNO is
operated by AURA, Inc. \ under contract to the National Science Foundation.}

\begin{abstract}
We present spectra of nine low-luminosity early-type galaxies 
(M$_{\rm B} > -17.5$) in the Coma
cluster.  The spectra, which were obtained with the Multiple Mirror Telescope
and cover the spectral region 3300-5350 \AA, 
exhibit a large variation in Balmer line strengths.  In fact the line
strength variation among the Coma low-luminosity galaxies is as
large as that found among a sample of low-luminosity ellipticals in the Virgo
cluster and 
lower density environments. When compared to detailed
population synthesis models, the
variation in Balmer line strengths among the Coma galaxies indicates a range in
luminosity-weighted mean age 
of from $\sim$1 Gyr to $\sim$12 Gyr.  The two youngest galaxies are shown to
be in a post-starburst state, i.e., they are not simply 
former spirals whose star formation was recently terminated.  Moreover, the
$\sim$1 Gyr ages of these two youngest low-luminosity galaxies are similar 
to those of the brighter post-starburst galaxies in Coma.

\end{abstract}

\keywords{ }

\section{Introduction}

Recent multi-fiber spectroscopy of the Coma cluster of galaxies has revealed
that a substantial fraction of the early-type galaxies in this nearby rich 
cluster have experienced a starburst within the last $\sim$1 Gyr (Caldwell 
et al.  1993, hereafter Paper I; Caldwell \& Rose 1997, hereafter Paper II).  
In particular, nearly 40\% of the early-type galaxies in the SW region of
Coma (centered $\sim$40\arcm SW of the cluster center) have experienced
recent starbursts, while in other areas of the cluster only $\sim$5\% of the
galaxies are in a post-starburst state\footnote[1]{There is some variation in 
what different investigators actually mean by the term ``starburst''.  For the
purposes of this paper we consider a starburst to be any period of unusually
high star formation for a galaxy of a given morphological type.  Hence for
{\it early-type} galaxies, any period of substantial star formation qualifies as
a starburst under our definition.}.

The aforementioned multi-fiber studies concentrated primarily on galaxies 
brighter than B = 17.5 (M$_{\rm B}$ = -17.5 for \hub = 70 \kms).  A substantial
number of galaxies with B$>$17.5 were also observed, but the signal-to-noise
ratio of these spectra was not high enough to make a reliable assessment as to
the presence or absence of recent starbursts.  Nevertheless, as is described in 
Paper I, a number of the faint galaxies (six out of 20 spectra in the SW
region) have spectra with an unusually weak break at Ca II H and K.  While the
spectra were generally too noisy for a reliable evaluation of the strength of the
Balmer lines in individual spectra, when the six weak-break spectra were
coadded, the resultant spectrum exhibited stronger Balmer lines and weaker 
metal lines than in the composite of ``normal'' break spectra (see Fig. 15 of
Paper I).  Hence on the basis of this previous spectroscopy, there is 
reason to suspect that some of the faint galaxies in Coma have also experienced
recent star formation, and perhaps starbursts.

Clearly it is important to determine whether the faint (B$>$17.5) galaxies in
Coma, especially those in the SW region, have participated in the starburst
phenomenon that many of their brighter, more massive brethren have
experienced.  In particular, a knowledge of the luminosity distribution of
the (post)-starburst galaxies in Coma could help to reveal the cause of the
starburst phenomenon, which is at present still highly uncertain.  Moreover,
it is important as well to determine whether the progenitors of the 
post-starburst galaxies were early-type galaxies, spirals, or may even be
spirals with recently truncated star formation (rather than true post-burst
galaxies).  Thus any information that can be gleaned concerning the nature
of the progenitor galaxies could also be useful in understanding the cause
of the starbursts.  With these questions in mind we have obtained spectra of
nine low-luminosity early-type galaxies in Coma with the MMT.  
The spectra exceed the S/N ratios of the multi-fiber spectra of the faint
galaxies in Papers 
I and II, and are of sufficient quality that age and metallicity
diagnostic indices developed in Rose (1994), Jones \& Worthey (1995),
Jones (1997), and Leonardi \& Rose (1996) can be applied with confidence.
In \S 2 the spectroscopic 
data are described.  In \S 3 we present results on the ages of the 
faint Coma galaxies, compare them to a sample of faint ellipticals in
Virgo and lower-density environments that have been studied by Jones (1997),
and finally investigate their kinematics within the Coma cluster.

\section{Observational Data}

\begin{figure}[p]
\plotone{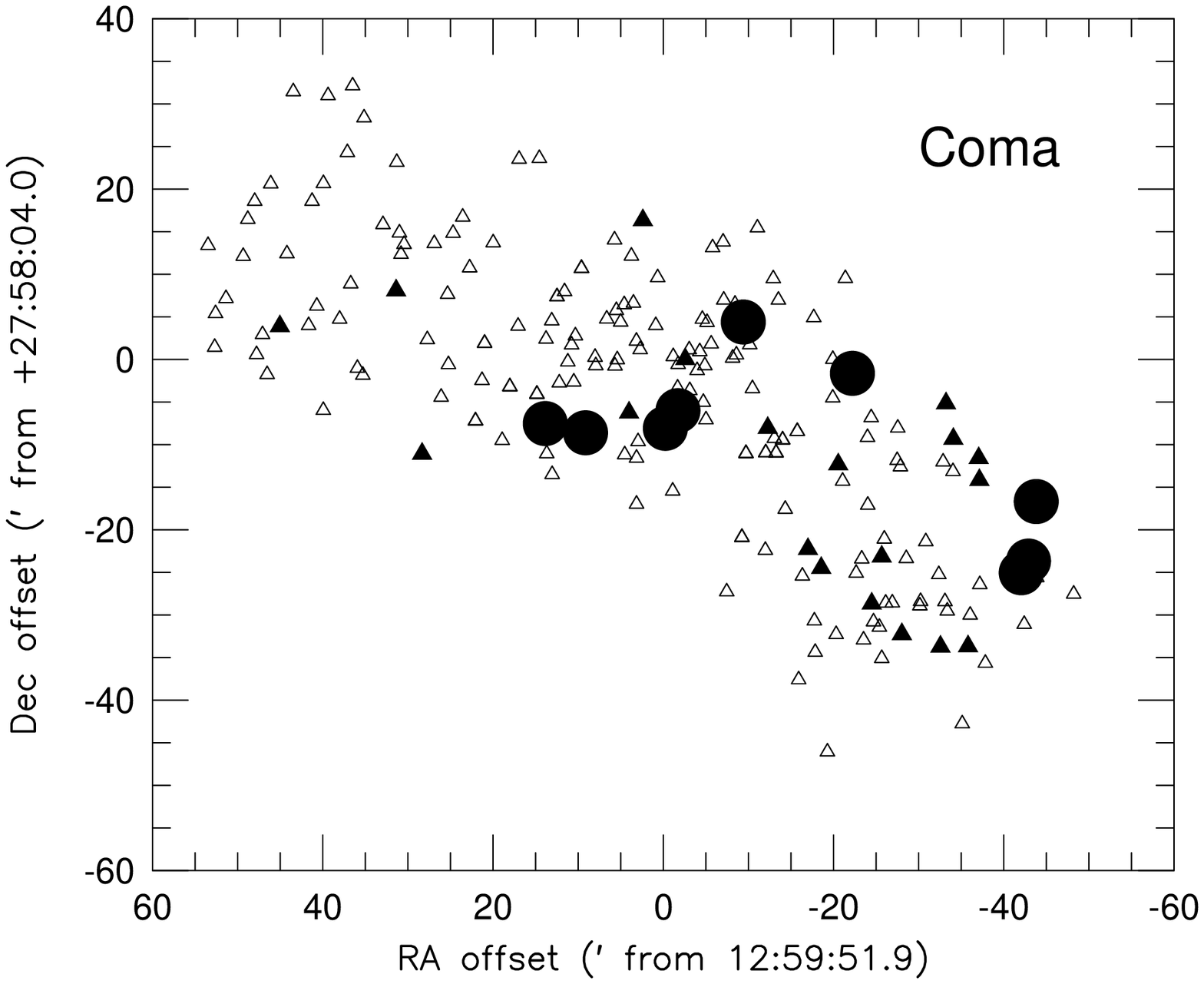}
\figcaption[caldwell.fig1.ps]{RA and Dec positions (measured in \arcm \ from the cluster center)
of the Coma cluster galaxies for which multi-fiber spectroscopy was acquired
in Papers I and  II.  Abnormal spectrum galaxies reported in those papers
are plotted as small filled
triangles.  The low-luminosity galaxies for which we have new MMT spectra are
plotted as large filled circles. \label{coma_lowl.pos}}
\end{figure}

We obtained spectra for nine low-luminosity galaxies in the Coma cluster.
The Coma galaxies selected are listed in Table \ref{tab1}. 
The first 5 are located in the central region, within 
15\arcmin \ (0.4 Mpc) of the cluster center. They had not been observed
before by us, but the velocities in Colless \& Dunn (1995) indicate that
they are cluster members.  The next 4 galaxies in the table are located
in the SW region.  All were observed previously by us in Paper I at lower 
S/N ratio; two were classified
as having weak H\& K breaks (GMP 5282 and 5320), one was thought to have
a normal break (GMP5362), and one was unclassifiable based on the 1993
data (GMP4420).  All of these galaxies are fainter than B=17.3, and all are
classified as S0's in Caldwell et al. (1993 and here), while one was
classified as an E by Dressler (1980).
In Fig. \ref{coma_lowl.pos} are plotted the locations in the Coma cluster
of the nine low-luminosity galaxies that we studied spectroscopically.

In addition, we observed one similarly low-luminosity galaxy in Virgo. 
This galaxy, IC 3065 (VCC140), was selected
from the Binggeli et al. (1985) catalog in the Virgo cluster; it 
has a low luminosity (B$_{\rm T}$=14.3,  B at Coma would be 18.1), and a 
velocity placing it in the Virgo cluster. It is 4.3 degrees from M87,
or about 1.1 Mpc.

\begin{figure}[p]
\plotone{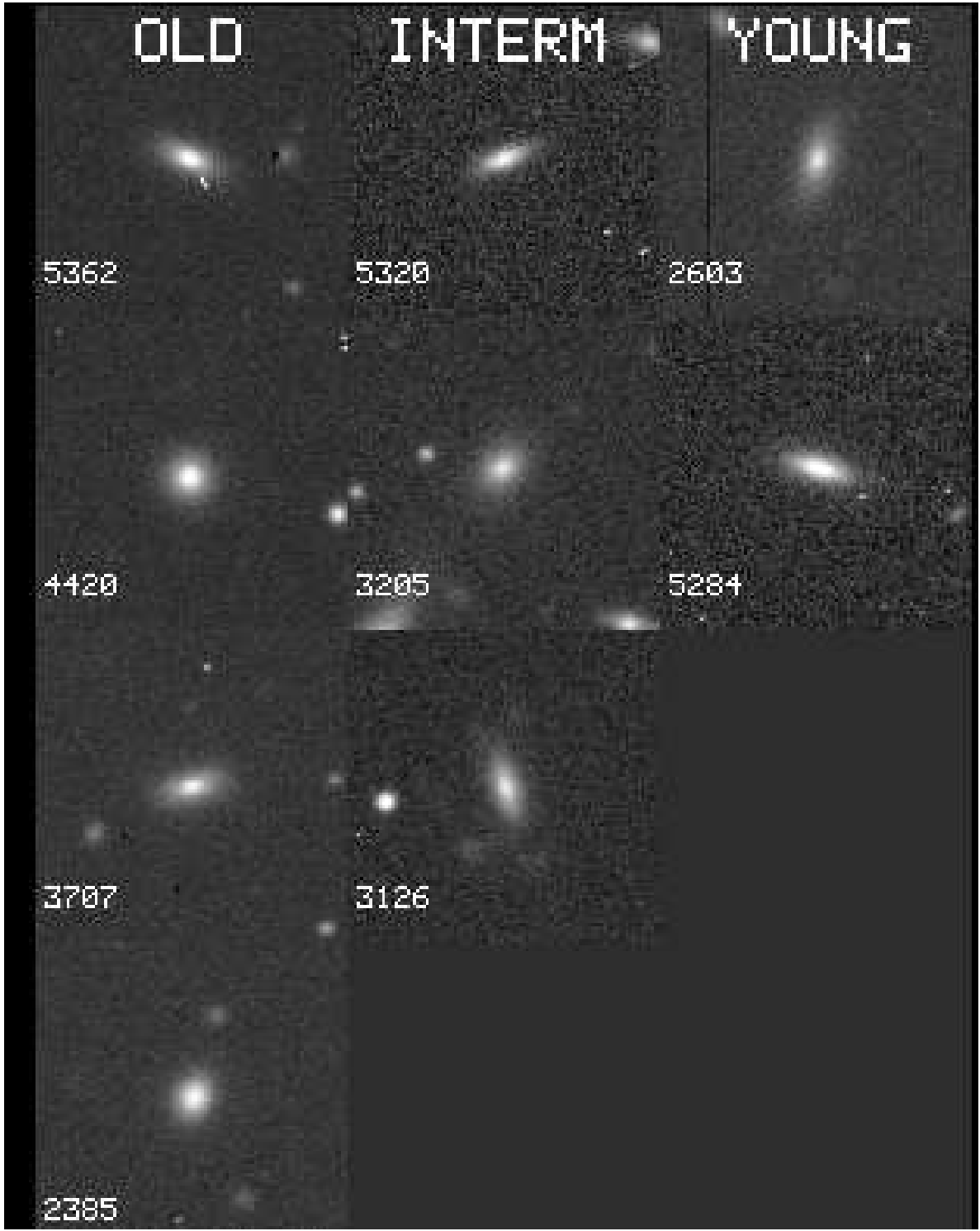}
\figcaption[caldwell.fig2.ps]{CCD Images of program galaxies, taken with the FLWO 1.2m telescope
using a V filter. The field size for each galaxy is about 1\arcmin.
Exposure times were 400 seconds \label{lowl_pic}}
\end{figure}

CCD images of the Coma galaxies were obtained at the FLWO 1.2 telescope
(Terlevich et al., 1997).  They are shown in Fig \ref{lowl_pic}, 
grouped into the age bins determined below.  Radial light profiles of these
galaxies confirm that all are S0's. 

\subsection{Spectroscopic Data}

Spectra of the 9 galaxies in Coma and of IC 3065 in Virgo were taken with the 
MMT and its blue-channel spectrograph on 1997 April 19-20.  The spectral 
resolution was 3.2\AA \ FWHM; exposure times are listed in Table \ref{tab1}.
In addition, spectra were obtained of the
radial velocity K giant standards HD171232, HD66141, and HD65934, and of the
spectrophotometric standard HZ44.  After the raw spectra were bias-subtracted 
and flat-fielded, they were reduced to one-dimensional, sky-subtracted form
using the IRAF apextract package, and then wavelength calibrated and fluxed.
Radial velocities were determined using the IRAF fxcor routine, and making
use of the radial velocity standard stars.
Velocity dispersions were also calculated from the IRAF fxcor program.  To do
so, however, required calibrating the broadening parameter determined in
fxcor so that a reliable transformation into a velocity dispersion could be
made.  Specifically, we
artificially broadened the spectrum of HD171232 by various amounts,
then cross-correlated the broadened spectra with the original spectrum,
and compared the broadening parameter produced by the fxcor routine with
the actual gaussian used to artificially broaden the spectra.

We were also kindly given spectra of 8 low-luminosity E galaxies in Virgo and
other lower-density environments by Lewis Jones, to form a comparison
sample with the Coma spectra.  These spectra, which were obtained at the KPNO 4-m
telescope at 1.8 \AA \ resolution FWHM, are summarized
in Jones's PhD thesis (Jones 1997).

\section{Ages, Metallicities, and Kinematics of Faint Coma Galaxies}

\subsection{Spectral Variations Among the Faint Coma Galaxies}

\begin{figure}[p]
\plotone{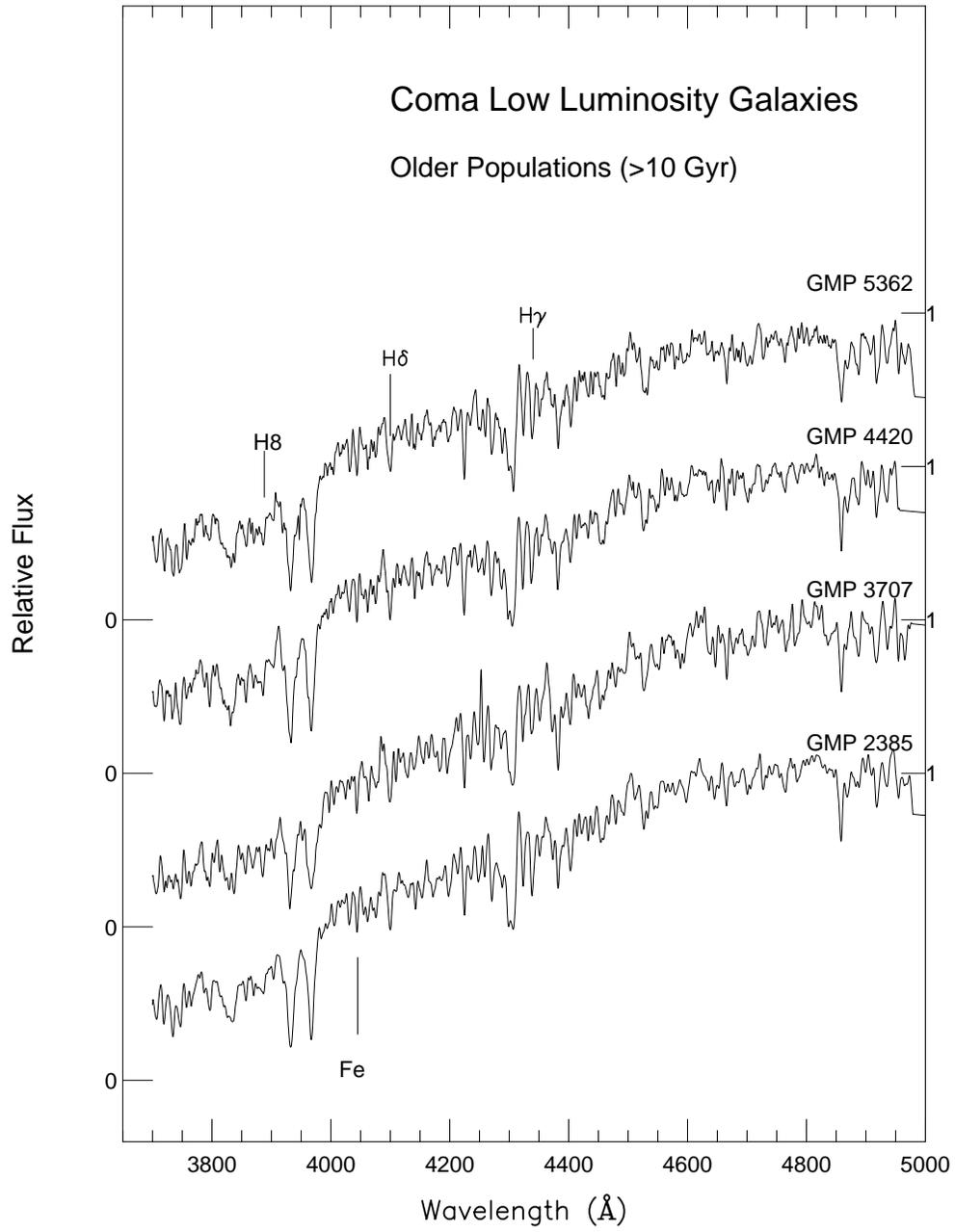}
\figcaption[caldwell.fig3a.ps]{(a) Spectra of ``old'' Coma cluster low-luminosity galaxies \label{lowl.spec}}
\end{figure}

In Fig. \ref{lowl.spec} are plotted spectra of the nine faint 
early-type galaxies in Coma and the one Virgo galaxy.  We have divided the
Coma spectra into three age groups, where the ages of the three 
groups will be determined in \S 3.2.
From an examination of the strength of the Balmer lines relative
to neighboring metal lines, it is apparent that there exists a considerable 
spread in the properties of the nine spectra.  We begin by quantifying this
variation in spectral properties.  

One manner in which to compare the strength of Balmer lines to neighboring
metal lines is to form a ratio of the central intensity of a Balmer line
relative to that of a nearby metal line (Rose 1984; Rose 1985).  Here we
compare H$\gamma$ to the Fe I$\lambda$4325 line, H$\delta$ to the 
Fe I$\lambda$4045 line, and H8 to the $\lambda$3859 feature (which is a mixture
of CN and Fe I absorption).  These ratios are hereafter referred to as the
\Hg, \Hd, and \H8 indices respectively, and are basically a measure of overall
spectral type.  Note that the way these indices are
defined, as the Balmer line becomes stronger relative to the metal line (i.e.,
towards earlier spectral type), the indices become lower in numerical value.
For instance, a star with spectral type A has an \Hd \ index of $\sim$0.5,
early G is at \Hd $\sim$0.8, and early K is at \Hd$\sim$1.2.  For this
paper we also define an average of the above three indices, which we 
hereafter refer to as the Hn/Fe index, i.e., Hn/Fe = $<$\Hg +\Hd +\H8$>$.
We will also make use of another line ratio index, defined in Rose (1985)
and modelled in Leonardi \& Rose (1996), which forms the ratio of the
combined feature Ca II H + H$\epsilon$ with the Ca II K line.  This
index, which is used to look at young populations due to its sensitivity
to the presence of hot stars, will henceforth be referred to as the Ca II index.
As discussed in Rose (1985), the line ratio indices have the advantage of
being relatively insensitive to variations in spectral resolution.

Observational errors in the line ratio indices have been determined from
multiple exposures of several galaxies in the following manner.  We obtained
three one hour exposures of GMP5320 and two each for GMP2603, GMP4420, and
GMP5362.  The rms scatter in the various indices were computed from each of
these multiple exposures, and the errors for individual galaxies were adjusted 
according to photon statistics to the flux level in the GMP4420 exposures.  It
was then assumed that the errors in \Hg, \Hd, and \H8 \ are the same, and so a
mean error for these three indices was computed from the average.  We then
compared this error with the error in the Hn/Fe I indices, which should be
$\sqrt{3}$ lower than the mean errors in the individual indices, according
to propagation of errors.  This turned out to be approximately the case, and
we adopted an error for Hn/Fe I of $\pm$0.03 for the flux level in GMP4420,
and adjusted the errors in all other galaxies for their different exposure
levels.  The errors in the individual indices were set to be $\sqrt{3}$
higher.

Data on the various Balmer line indices for the Coma low-luminosity galaxies
is summarized in Table \ref{tab2} and illustrated in 
Fig. \ref{hn}(a)-(c), where the three
individual indices (\Hg, \Hd, and \H8) are plotted against the composite
Hn/Fe index.  The Coma galaxies are plotted as filled circles, and the Virgo
low-luminosity galaxy IC3065 is plotted as a starred hexagon.  To avoid the
influence of velocity broadening of the lines on the spectral indices, all
spectra were gaussian smoothed to a common velocity dispersion of 100 \kms.
Also plotted, as unfilled squares, are the eight low-luminosity elliptical
galaxies from Jones (1997); these spectra were not only gaussian smoothed
to the same velocity broadening, but they were also broadened from their
instrumental resolution of 1.8 \AA \ FWHM to the FWHM of 3.2 \AA \ that is
characteristic of our Coma spectra.
From Fig. \ref{hn}(a)-(c) it is evident that all three Balmer indices are
well correlated, as expected, and demonstrate the quantitative range in Balmer
line strength that is apparent from visual inspection of 
Fig. \ref{lowl.spec}.  Furthermore,
the faint Coma galaxies and the low-luminosity Jones ellipticals clearly
exhibit similar behavior in these diagrams, i.e., the slopes and zeropoints
of fits to the Coma and to the Jones galaxies are indistinguishable within the
errors, and the range in index values is similar.  However, the Coma galaxies
are somewhat more concentrated to lower Balmer index values than the Jones
ellipticals, which we will see below is due to a younger mean age.

A more commonly used approach to defining spectral indices is to measure
the equivalent width of an individual spectral feature.  A problem with 
using such a procedure in the blue spectral region (at a spectral resolution
that is characteristically used for the study of galaxies in integrated
light) is that it is difficult to locate a clean continuum area on both sides
of the spectral line.  However, given the low velocity dispersion of the
Coma galaxies and the $\sim$3 \AA \ intrinsic resolution of the spectra, we
can use ``pseudo-equivalent width'' 
indices that have been defined previously
for such situations.  Specifically, Rose (1994) defined equivalent width
indices for H$\gamma$, Fe I $\lambda$4045, and Ca I $\lambda$4226, which were
revised by Jones \& Worthey (1995) and Jones (1997).  According to the Jones
(1997) definition, which we adopt here, the pseudo-continuum peaks on either
side of the line center are interpolated to find the local continuum at a given
point in the line; the equivalent width is obtained from the central 3.74
\AA \ around the line center (see Jones 1997 for details).  In what follows,
we make use of the \HgHR, \CaHR, and \FeHR \ indices from Jones (1997).
The great advantage to these indices is that Jones \& Worthey (1995) and
Jones (1997) have shown how they can be modelled in integrated light to
determine the ages and metallicities of galaxies without the usual degenerate
effects of age and metallicity.  The disadvantage is that the latter indices are
much more sensitive to the effects of spectral resolution (and differences in
spectrograph line profiles) than the line ratio indices.  Thus comparing indices
from two datasets (as we have done here) is less secure in the case of the
equivalent width indices than it is for the line ratio indices.  Data on the
equivalent width indices and their errors are also compiled in Table 
\ref {tab2}, where
the errors in indices were determined in the same basic manner as in the case
of the line ratio indices, i.e., from multiple exposures of several galaxies.

In Fig. \ref{hn}(d) we have plotted the \HgHR \ data for Coma and Jones galaxies versus
the Hn/Fe index.  Note that the agreement between the two datasets in this 
case is less striking than for the line ratio indices, namely, the Jones
galaxies show a shallower slope to their \HgHR \ versus Hn/Fe relation than
the Coma galaxies.   Plots of \FeHR \ and \CaHR \ versus Hn/Fe are shown in
Fig. \ref{agecal}.  In the case of \FeHR \ versus Hn/Fe, again the
Jones galaxies have a slightly shallower slope than the Coma galaxies.  
The differences between Jones's (1997) galaxies and ours in the
pseudo-equivalent width diagrams is likely due to the above-mentioned
difficulty of landing such indices from different spectrographs onto a
common system (without the use of ``standard'' stars to supply transformations
between data sets), due to the sensitivity of the indices to the spectrograph
line profiles as well as to the overall FWHM.  Since we do not have observations
of ``standard'' stars from the Jones (1997) stellar library to tie our
system to his, we are unable to rule out the possibility that the
pseudo-equivalent width index differences reflect intrinsic differences
between our Coma galaxies and his galaxy sample.
Finally, the
most noteworthy aspect of the \CaHR \ versus Hn/Fe plot is the relatively 
large scatter in both sets of galaxies.  This lack of correlation between Ca 
and overall spectral type is discussed in Jones (1997) and found to be part
of a general pattern of non-solar abundance trends in $\alpha$-elements.
Such non-solar abundance behavior has been advocated in several previous studies
(e.g., Faber 1977, Worthey et al. 1992, de Freitas Pacheco 1996, Casuso et al.
1996), especially in regard to Mg, and is attributed to differing chemical
evolution histories among early-type galaxies.

To summarize at this point, the faint Coma early-type galaxies exhibit a
quantifiable spread in their Balmer line properties.  This spread is very
similar in nature to that found in Jones's (1997) sample of low-luminosity
elliptical galaxies.  The similarity between the Coma and Jones galaxies is
particularly apparent in the line ratio indices, which are relatively
insensitive to spectral resolution effects.  The agreement is less compelling
in the case of the equivalent width \HgHR \ and \CaHR \ indices, but we
attribute this to the difficulty of landing two different data sets on the
same system.  Finally, the \CaHR \ indices indicates a sizable spread in
non-solar abundance ratios among the faint Coma galaxies that is similar
to that found in Jones's low-luminosity ellipticals.  We now consider how these
spectral variations can be modelled in terms of age and metallicity
differences.

\subsection{Age Variations Among the Faint Coma Galaxies}

The most significant challenge to a reliable determination of the ages and
metallities of galaxies from their integrated spectra is the degenerate
effects of age and metallicity on the integrated spectra (e.g., Worthey 1994).  
Recently, however,
Jones \& Worthey (1995) and Jones (1997) have used the \HgHR \ index, plotted
versus a number of metal line indices, to produce age determinations that are
relatively free of the age-metallicity degeneracy.  The age and metallicity
determinations are based on population synthesis models described in 
Worthey (1994). In these models, stellar evolutionary isochrones from 
several sources are combined with empirical fits to the behavior of spectral
indices in stars as a function of T$_{eff}$, log g, and [Fe/H] to predict
the behavior of integrated spectral indices as a function of age and
metallicity.  The application of this modelling to elliptical galaxies is
described in Jones \& Worthey (1995) and Jones (1997).  In particular, Jones
(1997) finds that among the eight low-luminosity ellipticals he studied in
Virgo and in lower-density environments, there is a large range in ``age''
(i.e., epoch of last major star formation) among the ellipticals, but very
little dispersion in metallicity.  Thus he finds age to be the principal 
contributor to the observed spread in spectral properties of low-luminosity
elliptical galaxies.  Ages of from 2 to 16 Gyr were found among the eight
galaxies studied.

As was illustrated in \S3.1, the spread in Balmer line spectral indices is as
great among the faint Coma early-type galaxies as it is among Jones's (1997)
low-luminosity ellipticals.  It is thus natural to suspect that there is a
correspondingly large range in age among the faint Coma galaxies.  Some
support for this conjecture is evident in Fig. \ref{mod}, where the \HgHR \ indices
for the Coma (filled circles) and Jones (unfilled squares) galaxies are plotted
versus the \FeHR index.
This diagram separates age from metallicity effects, with the Jones galaxies
roughly following a line of constant (near-solar) metallicity and varying age
from the lower right (older age) to upper left (younger age) area of the
diagram.  The faint Coma galaxies follow the same basic trend as the Jones
galaxies, but the errors in the spectral indices are considerably larger.  Thus
the age sensitive \HgHR \ versus \FeHR \ plot indicates a large spread in age
among the faint Coma galaxies, but the errrors are large enough that the ages
derived from this plot are probably not very accurate on an individual basis.
In addition, as pointed out earlier, the \HgHR \ and other
pseudo-equivalent width indices measured for the Coma galaxies do not appear
to exactly match the Jones (1997) system, thereby casting some doubt on the
reliability of the ages that we would infer from these indices.

To obtain a more accurate and reliable determination of the ages of the faint Coma galaxies,
we make use of the fact that Jones (1997) found a good correlation between 
the \HgHR \ indices (from which the ages are determined) for his eight 
galaxies and the \Hd \ line ratio indices, which in turn correlate well with 
the other two Balmer line ratio indices (\Hg \ and \H8).  To illustrate this
point we plot in Fig. \ref{agecal} the \HgHR \ indices for Jones's (1997) eight galaxies
versus the combined Hn/Fe indices.  In this plot we first smoothed Jones's
spectra to the same spectral resolution as our slightly lower resolution
MMT Coma spectra.  It is evident from Fig. \ref{agecal} that \HgHR \ correlates well
with Hn/Fe.  Thus we now use Hn/Fe as a surrogate for \HgHR, since, as a
composite of three indices, it has smaller errors than \HgHR.  To do so, we
made a fit of the Hn/Fe indices of Jones's eight galaxies to the ages he
derived for them in Jones (1997) by modelling their \HgHR \ indices.  
We then used the fit to
convert the Hn/Fe indices of the faint Coma galaxies (and IC3065) into ages.  
The results of this procedure are listed in Table \ref {tab3}.  
Note that two of the
faint Coma galaxies have a young ($\sim$2 Gyr) age, three fall in an
intermediate ($\sim$4-6 Gyr) age range, and the other four are basically
old ($\sim$10-12 Gyr).  The age range is similar to that of Jones's (1997)
eight low-luminosity ellipticals, except that none of the faint Coma galaxies
appears to be as old as Jones's two oldest galaxies ($\sim$16-17 Gyr), and
on the whole the Coma galaxies are somewhat younger than Jones's 
(only one of which
is young, one is intermediate age, and the other five are old).  Again,
we note that there is some uncertainty to using the Hn/Fe index as a
surrogate for \HgHR , but the tight correlation between the Hn/Fe and
\HgHR \ indices of the Jones (1997) galaxies, coupled with the fact that
our galaxies appear to exhibit the same behavior in the Fig. \ref{mod}
diagram of \HgHR \ versus \FeHR , indicates that the substitution is valid.

For the two youngest Coma galaxies, with determined ages of $\sim$2 Gyr,
it is of course highly likely that we are observing recent star formation
superimposed on an older stellar population(s), rather than a
uniformly $\sim$2 Gyr old population.  To evaluate the possible star formation
scenarios for these two ``youngest'' Coma galaxies, we utilize another
age determination technique for young stellar populations that is described
in Leonardi \& Rose (1996).  The Leonardi \& Rose (1996) method, which is 
based on the above-mentioned Ca II index, is geared towards distinguishing 
between  different possible star formation scenarios for a galaxy with recent 
star formation.  For example, if recent star formation (i.e., within the last 
$\sim$2 Gyr) has occurred, how do we determine whether the star formation 
occurred in a relatively sudden burst, or whether we are simply
viewing a galaxy at some point after truncation of long-term star formation?
And if we are viewing the galaxy after a star formation burst, how can we
distinguish between a weak burst, seen relatively soon after it terminated, 
from a stronger burst seen relatively longer after termination?  For details of
the method, the reader is referred to Leonardi \& Rose (1996). The
method has been applied to several of the brighter Coma early-type galaxies
in Caldwell et al. (1996).  There it was found that the brighter Coma 
galaxies did experience recent starbursts, as opposed to truncation of
long-term star formation (e.g., in the disk of a spiral).  

In Fig. \ref{agedate} we
reproduce the basic Leonardi \& Rose (1996) diagram with the positions of
the two ``young'' faint Coma galaxies (GMP2603 and GMP5284) marked as filled 
triangles.  Plotted as unfilled squares are post-starburst models shown at
various times after termination of the starburst; the specific times are
marked on the plot (e.g., 0.0., 0.5, 1.0, etc. Gyr).  The filled circle 
represents the composite of five of Jones's (1997) old elliptical galaxies.
\footnote[2]{Note that we have used a different composite old population
spectrum than in Leonardi \& Rose (1996), where we used a composite spectrum 
of 70 Coma cluster galaxies obtained with the KPNO 4m Hydra multi-fiber
positioner.  The composite of low-luminosity elliptical galaxy spectra used
here seems more appropriate to us for comparison to the faint Coma galaxies,
which have similar luminosities, but the results obtained differ by less than
0.1 Gyr if the composite Coma cluster galaxy spectrum is used as the old
population anchor.}
The dotted lines that connect the various post-starburst points with the
composite old galaxy point represent the results of taking linear combinations
of the former with the latter.  The small x's on the dotted lines represent 
ten percent increments in the balance between PSB and old galaxy light.  Thus
the first x next to the 0.0 age PSB point represents a combination of 90\%
(normalized at 4000 \AA) PSB light and 10\% old galaxy light.  Finally, we have
plotted the trajectory of a ``truncated'' galaxy which had a constant star
formation rate over 15 Gyr, and then the star formation was suddenly
terminated.  The various open triangles connected by the solid line represent
the spectral indices of this fading truncated galaxy at increments of 0.5 Gyr
after truncation of the star formation, with the first triangle near the
``CON'' label representing the galaxy 0.0 Gyr after truncation.  Thus a galaxy
with such a truncated star formation history should lie somewhere on the solid
line, while a truncated galaxy that had a declining star formation rate over
its lifetime would lie somewhere to the right of the line.  Given that the
two faint Coma galaxies lie to the left of the truncated constant star
formation line, we conclude that they underwent a true starburst in
the sense that they experienced enhanced star formation before going
quiescent.  Furthermore, as can be seen from Fig. \ref{agedate}, the PSB age 
of GMP5284, which lies in the SW region of Coma populated by many other PSB
galaxies, is $\sim$1.0 Gyr, which is consistent with the $\sim$0.8-1.3 Gyr
ages found for the four brighter Coma PSB galaxies in the SW region discussed 
in Caldwell et al.  (1996).  For reference, the positions of three of those 
galaxies are marked as
filled squares in  Fig. \ref{agedate} (the fourth galaxy was observed at lower
spectral resolution and thus cannot be marked on the plot).  Finally, the age
of GMP2603, which is the low-luminosity ``young'' galaxy in the central region
of Coma, can be seen in Fig. \ref{agedate} to have a PSB age of $\sim$1.6 Gyr,
which is also about the same age as that of the PSB galaxies in the SW region.

To conclude, the two ``youngest'' low-luminosity galaxies studied by us
(GMP2603 and GMP5284) both
are seen in Fig. \ref{agedate} to have undergone recent starbursts, i.e., we
are viewing a recent episode of star formation in these two galaxies that is
superimposed on an older population.  Their spectral indices do {\it not} 
appear to be consistent with a ``truncated spiral'', i.e., a constant star 
formation spiral whose star formation was truncated in the recent past.  
Like the more luminous PSB Coma galaxies, there is little 
evidence for a disturbed morphology in the images of these two galaxies 
(see Fig. \ref{lowl_pic}). Thus
these two faint early-type galaxies appear to have undergone a similar star
formation history in the recent past as the brighter PSB galaxies in the Coma
cluster.  

\subsection{Spatial Distribution and Kinematics of the Faint Coma Galaxies}

In the previous section we showed that the ages of the faint Coma early-type
galaxies obtained from modelling of high S/N ratio MMT spectra correlate
well with the separation of the multi-fiber spectra from
Paper I into two groups (based on visual examination of the spectra),
those with weak Ca II breaks versus those with normal break amplitudes.  
Specifically, the modelling has determined that the
group of faint galaxies with weak break amplitude spectra consists of
galaxies which have undergone major star formation within the last $\sim$2 Gyr.
We now consider the spatial distribution and kinematics of the low-luminosity
Coma galaxies in the SW that were visually classified as weak break, and compare
to those of their normal break amplitude faint brethren.  Basic information
on the weak break and normal break faint galaxies is given in Tables 
\ref {tab4} and \ref {tab5},
respectively.  There the galaxies are listed by GMP number and position and
velocity data are also given.

In Fig. \ref{coma_lowl.xy} we have plotted the spatial distribution of the weak break (filled
squares) and normal break (open squares) faint galaxies in Coma.
There appears to be no spatial
segregation of the weak-break low luminosity galaxies
from the normal break ones, though the small number of galaxies involved
here would make such a distribution hard to detect.

On the other hand, the kinematics of the two groups of faint galaxies in the
SW region of Coma appear to be different.  The mean radial velocity
and line-of-sight velocity dispersion for the fifteen normal break galaxies are
6845 \kms and 985 \kms respectively, while the mean velocity and dispersion
for the six weak break galaxies are 7505 \kms and 137 \kms respectively.  
The hypothesis that the
two groups are selected from the same parent velocity distribution is
rejected by the Kolmogorov-Smirnoff test at the 4.4\% level.
Thus while the evidence is not overwhelming,
there is an indication from the somewhat limited velocity data that the
faint Coma galaxies in the SW with recent star formation belong to a higher
mean velocity and lower velocity dispersion system than the ``older'' faint
galaxies.  In comprehensive analyses of the kinematics of the Coma cluster,
both Colless \& Dunn (1996) and Biviano et al. (1996) found a gradient in
velocity along the NE-SW axis of the cluster.  Biviano et al. (1996) in
particular found that if the sample is restricted to faint (i.e., 17$<$B$<$20)
members, then the mean velocity in the SW region of Coma is $\sim$6600 \kms;
they demonstrate that the faint galaxies tend to trace the underlying kinematics
of the cluster while the brighter galaxies tend to lie in subclusters.
On the other hand, there is a spatial-velocity subcluster in the SW region,
centered on NGC4839, which has a high mean velocity ($\sim$7339 \kms, according
to Colles \& Dunn 1996), and low dispersion ($\sim$329 \kms).  Thus according
to the evidence accumulated so far, the normal break faint Coma galaxies in
the SW region appear to be partaking in the overall kinematics of the faint
galaxies in that region, while the weak break galaxies, i.e., those with
recent star formation, appear to be affiliated kinematically with the
NGC4839 group.  To place this in further perspective,
Biviano et al. (1996) find that the abnormal spectrum galaxies from Paper
I have a mean velocity and dispersion very similar to that of the NGC4839
subcluster.  Thus the brighter post-starburst galaxies in Coma SW appear to
be associated with the NGC4839 kinematic subcluster. However, the velocity
dispersion of the abnormal spectrum galaxies remains controversial in that
in Paper I we found a high value, while Biviano et al. (1996), using the
biweight estimator, find a low value.  In short, while the data is still
sparse, the evidence to date sugests that the fainter galaxies with recent
star formation belong to the NGC4839 subcluster, along with the brighter
post-starburst galaxies.  There remains considerable uncertainty as to whether
the NGC4839 subcluster is in the process of falling into the main cluster
(Colless \& Dunn 1996) or has already passed through
the main cluster and is in the process of merging with it (Burns et al. 1994;
Paper II).

\section{Conclusions}

The two main results to emerge from spectroscopy of the nine low-luminosity
early-type galaxies in Coma are:\\
(1) The faint Coma galaxies have a large range in star formation histories.
In fact the variation in age found among the faint Coma galaxies is at least
as large as, and perhaps even larger than, that found among low-luminosity
ellipticals in lower density environments.  \\
(2) We have shown that two of the nine galaxies contain a very recent ($<$2 Gyr)
population superimposed on an older population.  In both cases the $<$2 Gyr
population represents a period of enhanced star formation, i.e., we are seeing
both galaxies in a true PSB state, rather than in a state of recently truncated
star formation in a disk.  Thus these two galaxies appear to have had similar
recent star formation histories as the brighter PSB galaxies that are 
concentrated in the SW region of Coma.

The fact that the faint Coma cluster S0 galaxies have such a large
range in ages provides an interesting contrast to studies of the brighter
early-type galaxies in rich clusters.  In the central regions of Coma, most of
the brighter galaxies have uniform (old) ages (Bower et al. 1990; Bower et al.
1992; Rose at al. 1994), while the age range among brighter early-type
galaxies in the SW region is large.
Thus the evolution of cluster galaxies appears to be a 
function of mass as well as of location in the cluster.
Recent HST images have revealed that over time, disk galaxies in clusters 
evolved from star-forming systems to non-star-forming 
systems (e.g., Dressler et al. 1994).  Our results indicate that the low
luminosity disk galaxies in Coma evolved to the latter state on a longer
time scale (on average) than the more luminous ones.  Simulations of disk 
galaxies in
clusters, such as those of Moore et al (1996, 1997) which follow the 
transformation of a small
disk galaxy into a spheroidal system, may be relevant here. However,
what is needed is not necessarily a drastic structural change (the faint Coma
galaxies studied by us are S0's, rather than spheroidals), but rather
a method to remove the gas via vigorous star formation over a short time period.

\acknowledgements

We wish to thank Lewis Jones for providing his spectra of eight low-luminosity
elliptical galaxies.
The observations reported here were obtained with
the Multiple Mirror Telescope, which is jointly run by the
Smithsonian Institution and the University of Arizona.  This research was
partially supported by NSF grant AST-9320723 to the University of North
Carolina.

\clearpage

%
%



%

\newpage

\clearpage
\clearpage

\clearpage
\begin{figure}[p]
\plotone{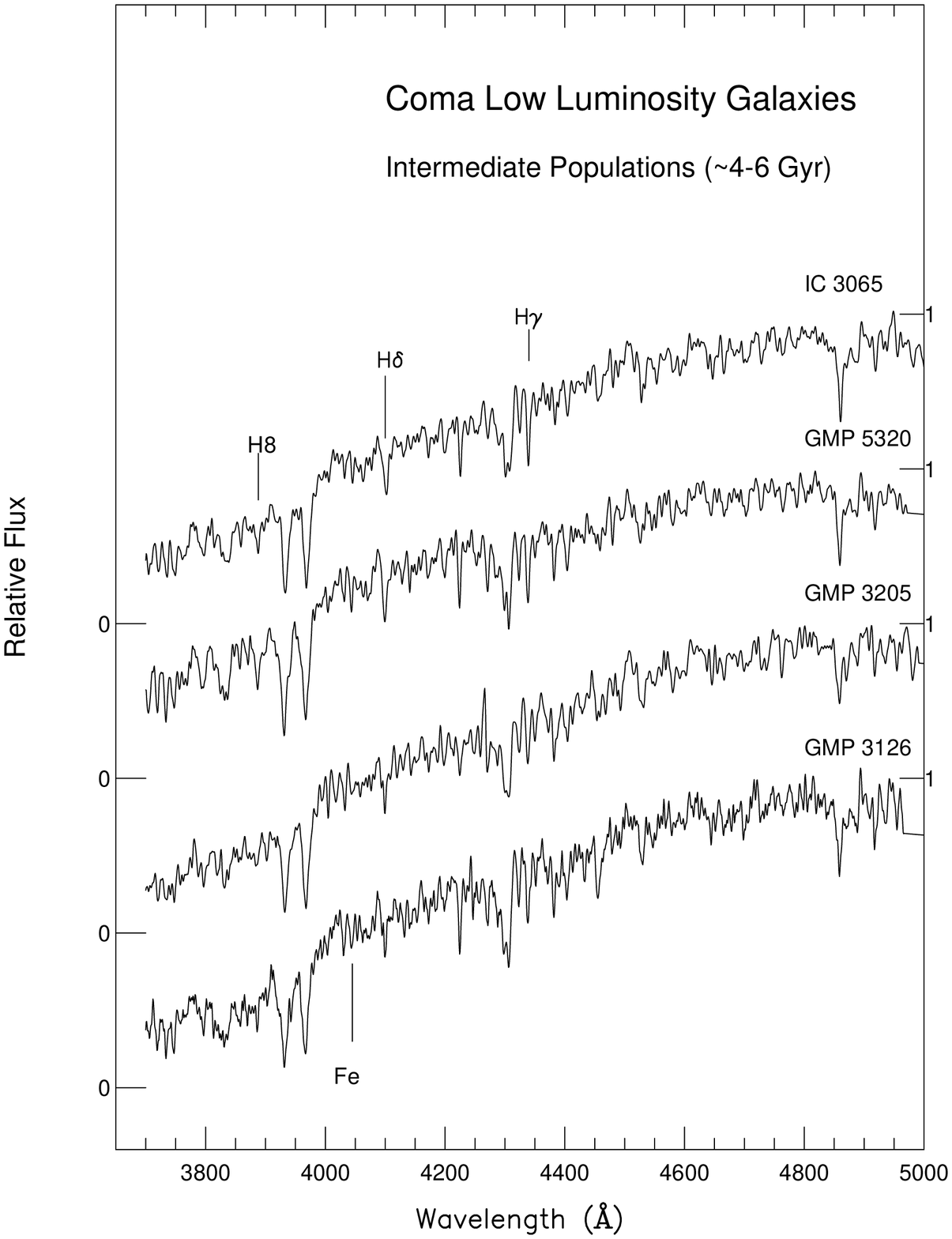}
\figurename{ 3(b) Spectra of ``intermediate'' age Coma cluster low-luminosity galaxies,
and a similar galaxy in the Virgo cluster (IC3065). \label{lowl.spec2}}
\end{figure}
\clearpage
\begin{figure}[p]
\plotone{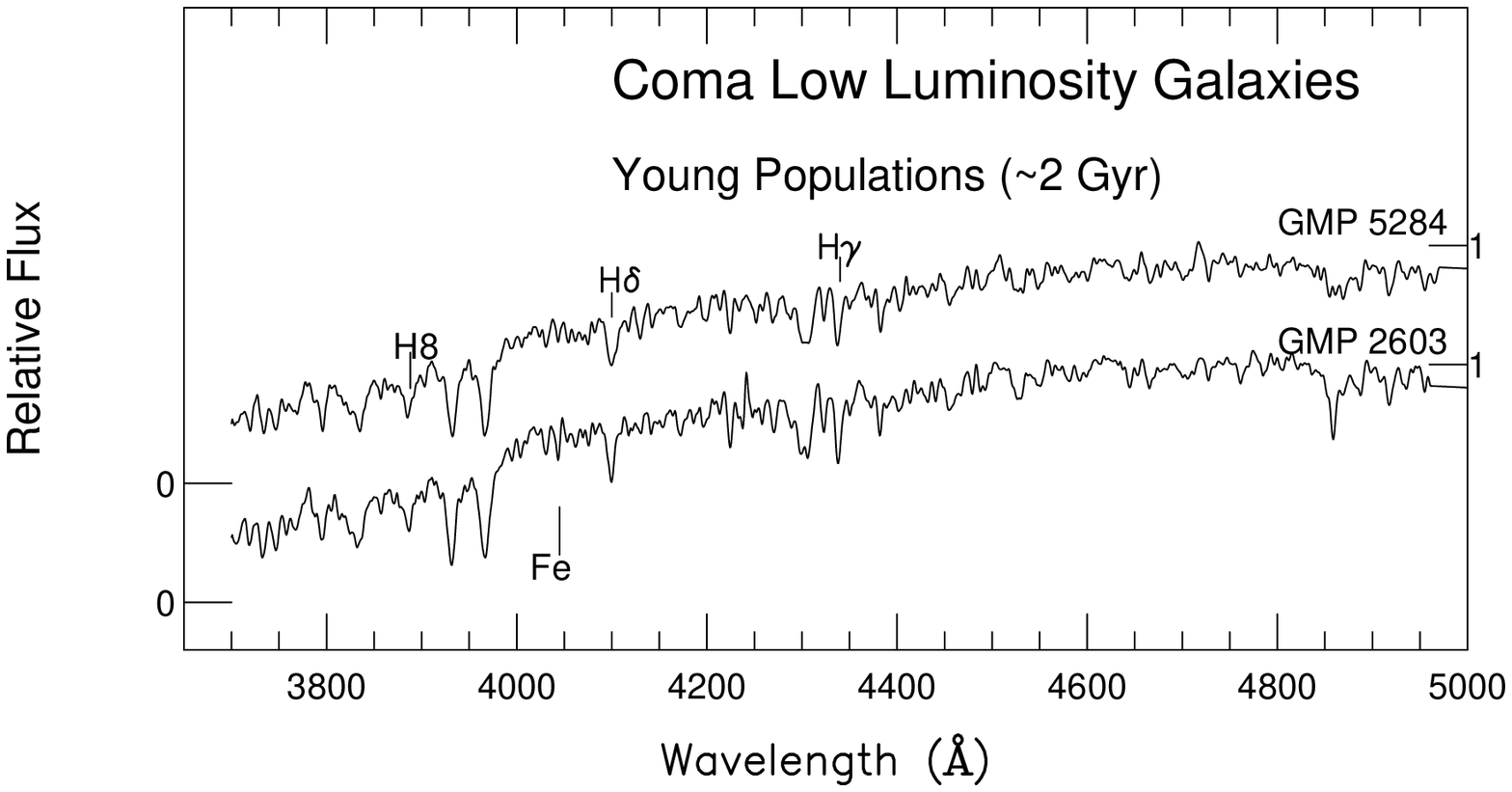}
\figurename{ 3(c) Spectra of ``young'' Coma cluster low-luminosity gaalxies.\label {lowl.spec3}}
\end{figure}
\clearpage
\begin{figure}[p]
\plotone{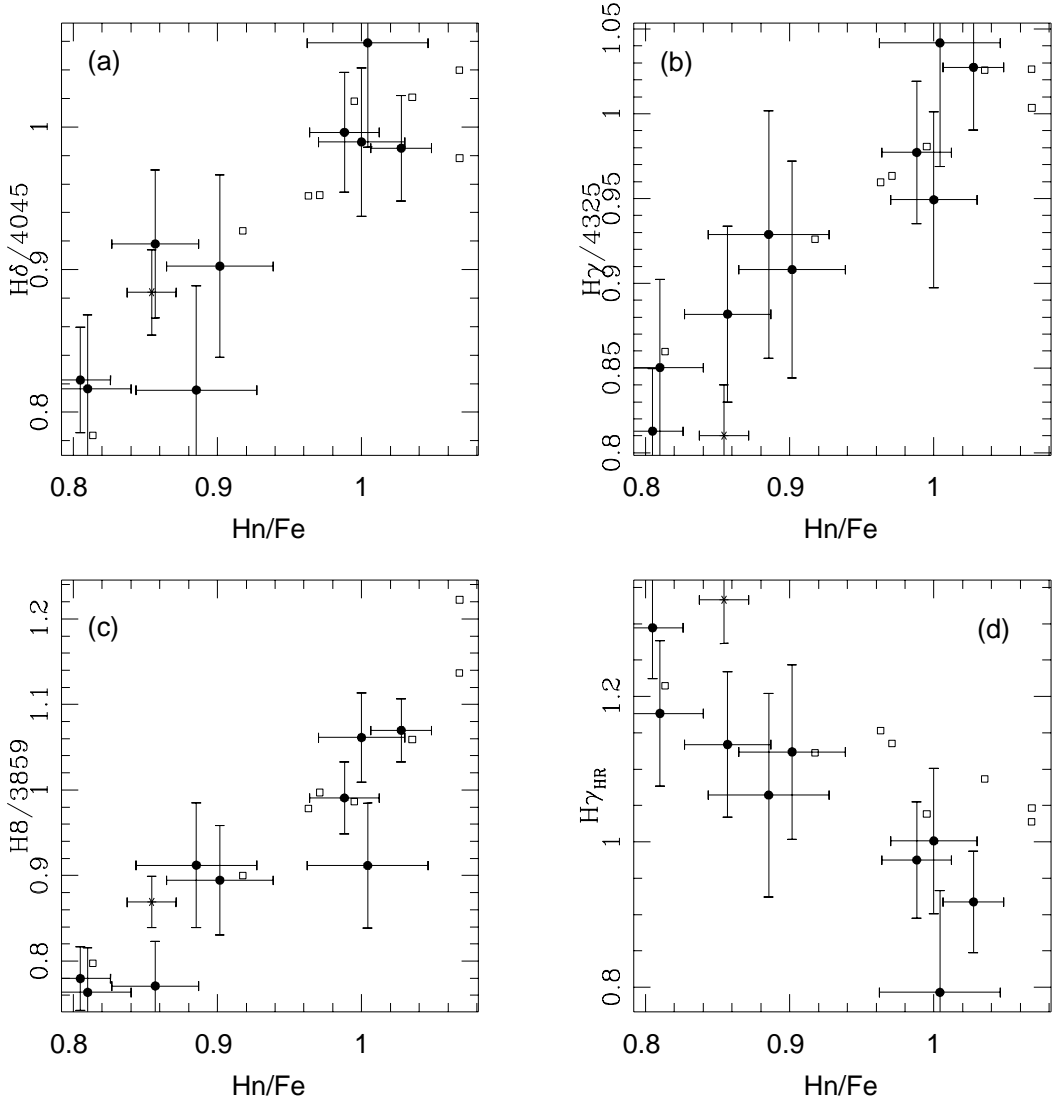}
\figcaption[caldwell.fig4.ps]{Individual Balmer line ratio indices are plotted versus the composite
Balmer line ratio index Hn/Fe in (a) \Hd, (b) \Hg, and (c) \H8.  In (d) the 
H$\gamma$ equivalent width index \HgHR \ is plotted.  The faint Coma galaxies
are represented as filled circles, IC3065 is represented as an asterisk, and
the Jones (1997) low-luminosity ellipticals are plotted as open squares.  Error
bars are $\pm$1 $\sigma$ \label{hn}}
\end{figure}
\clearpage

\begin{figure}[p]
\plotone{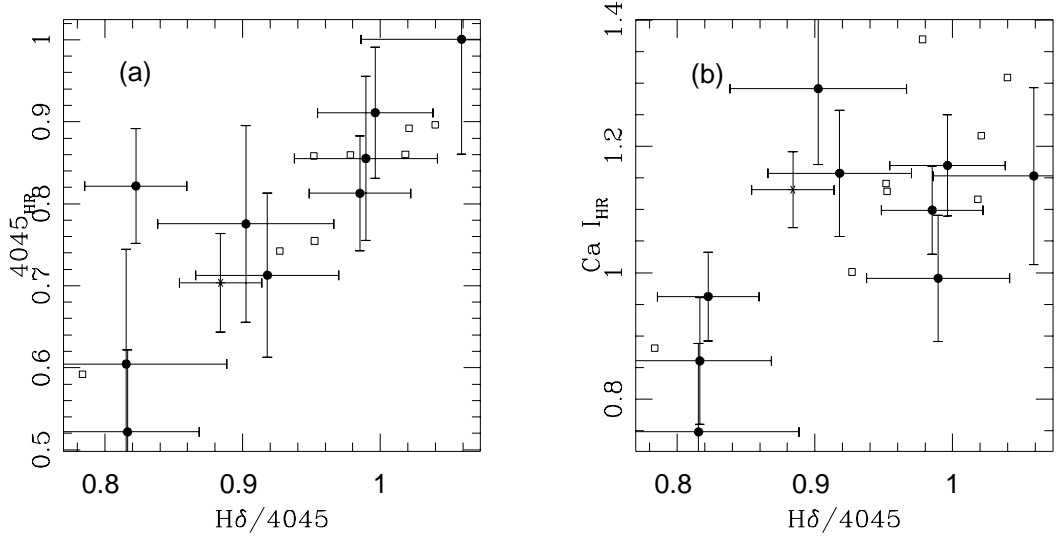}
\figcaption[caldwell.fig5.ps]{Equivalent width indices \FeHR (a) and \CaHR (b)  are plotted 
versus the composite Balmer line ratio index Hn/Fe. Same symbols as in Fig. 4 \label{eqw}}
\end{figure}
\clearpage
\begin{figure}[p]
\plotone{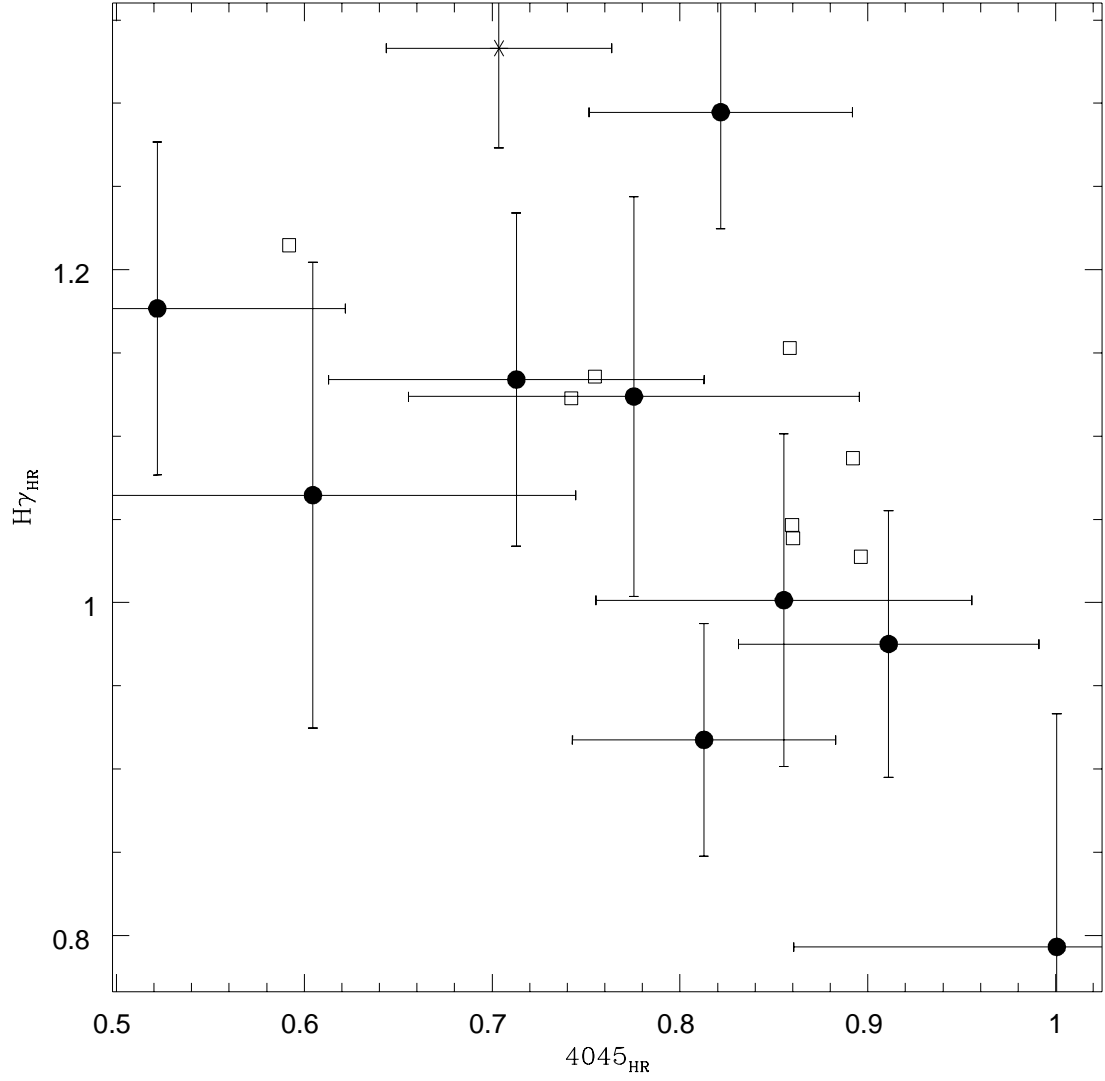}
\figcaption[caldwell.fig6.ps]{The \HgHR \ index is plotted versus the \FeHR \ index.  Same symbols 
as in Fig. 4 \label{mod}}
\end{figure}

\clearpage
\begin{figure}[p]
\plotone{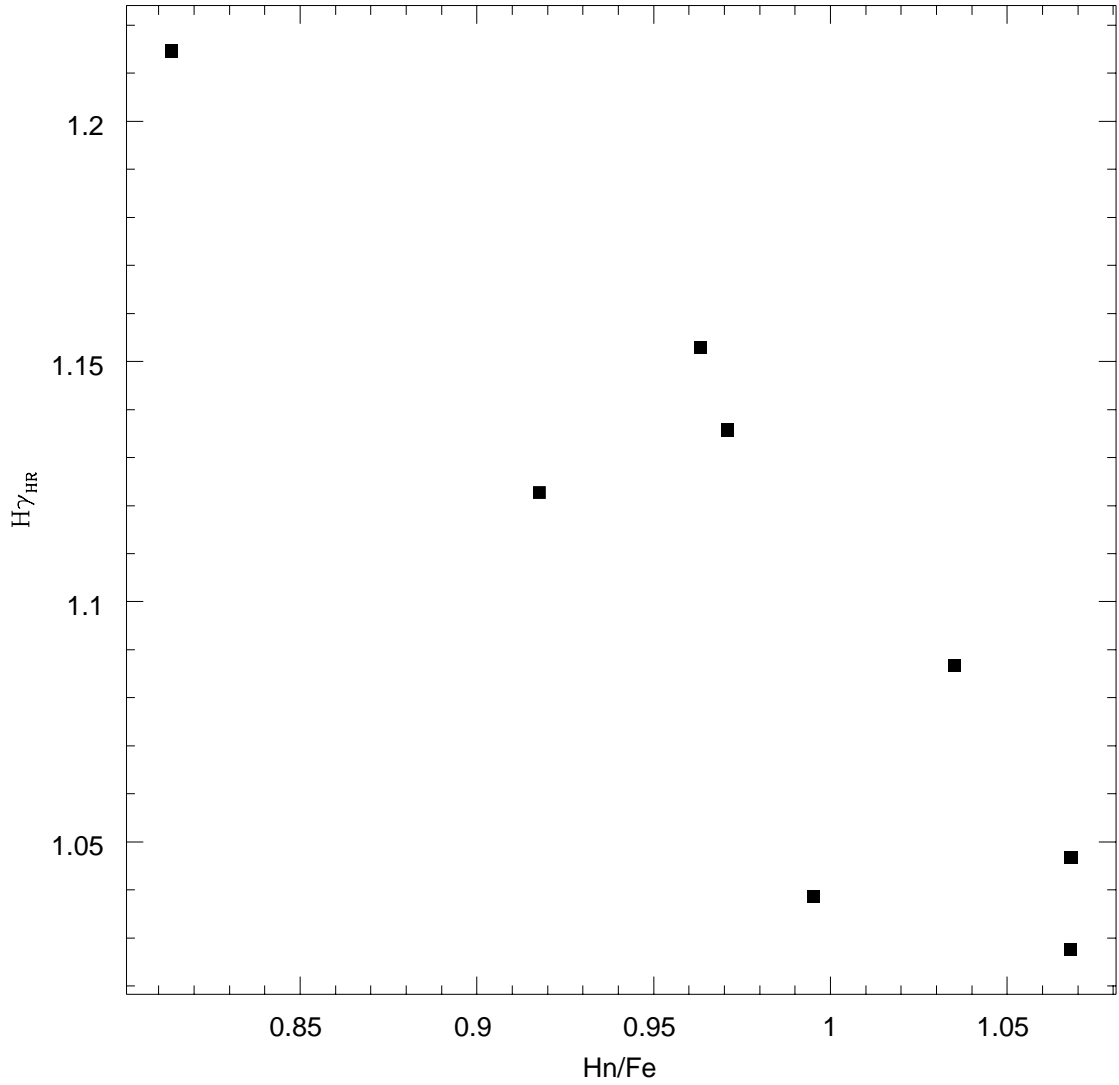}
\figcaption[caldwell.fig7.ps]{The age sensitive \HgHR \ index is plotted versus the Hn/Fe index for
Jones's (1997) eight low-luminosity ellipticals.  The good correlation between
the two indices indicates that Hn/Fe can be used as a surrogate for \HgHR \ in
age determinations \label{agecal}}
\end{figure}
\clearpage

\begin{figure}[p]
\plotone{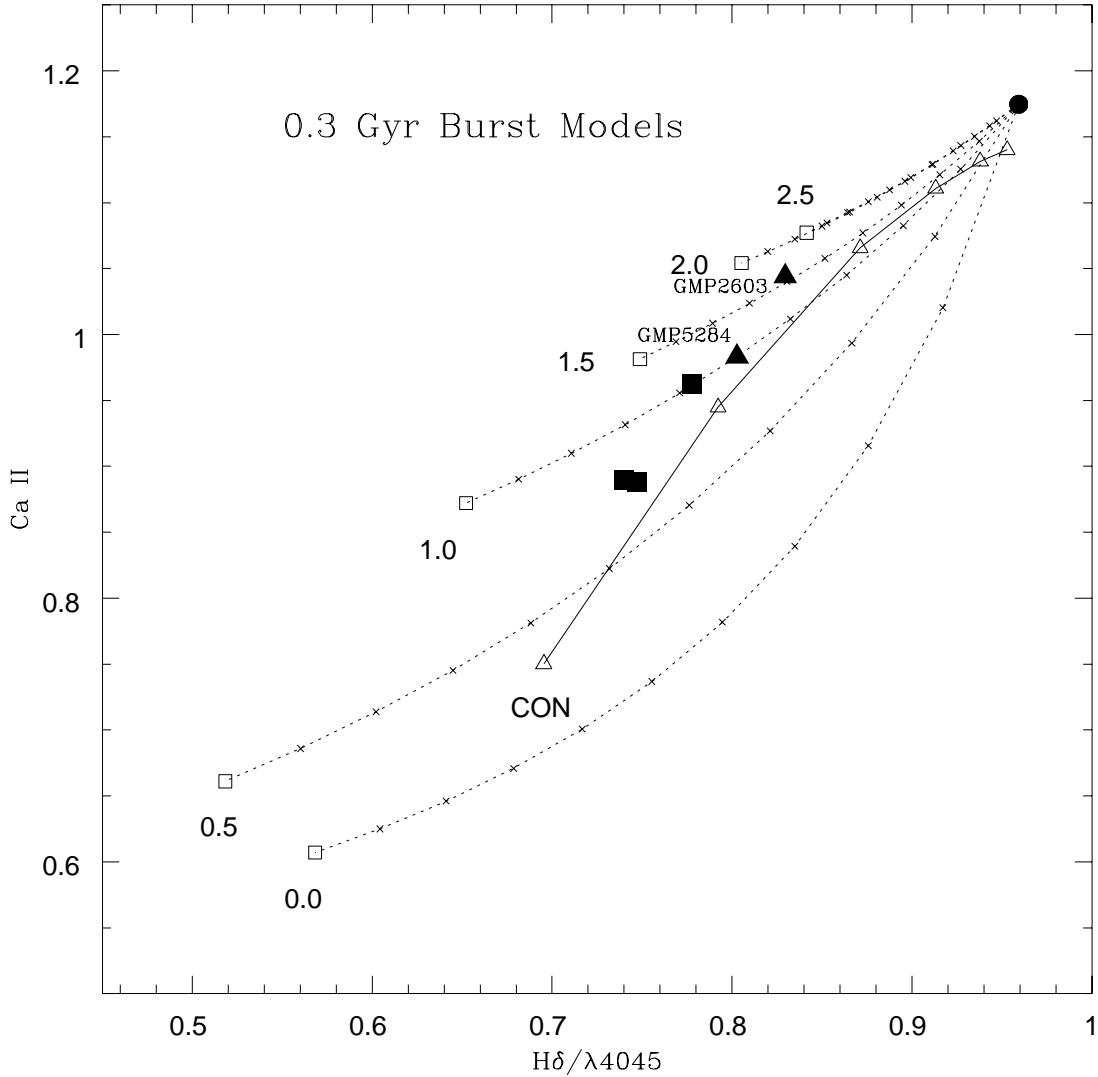}
\figcaption[caldwell.fig8.ps]{Ca II versus \Hd \ diagram for burst models and for various Coma 
galaxies.  Curved, dashed lines are derived from linear combinations of
post-starburst model spectra with the observed composite spectrum of 
an old galaxy population.
Unfilled squares represent Bruzual and Charlot (1995) model spectra
for a pure 0.3 Gyr long starburst that is seen at the noted times after
termination of the burst.  The small crosses designate 10\% increments in the
balance of burst versus old population light, normalized at 4000 \AA.  The 
solid line marked ``CON'' represents the evolutionary track of a truncated
constant (over 15 Gyr) star formation population seen at different times after
termination of star formation.  The unfilled triangle at the bottom of the line
represents the index values right after the truncation of star formation, while
each successive triangle denotes a time step of 0.5 Gyr.  Also plotted are the
two faint Coma galaxies GMP2903 and GMP5284 (filled triangles), and three
PSB galaxies (filled squares) previously studied in Caldwell et al. (1996) \label{agedate}}
\end{figure}
\clearpage

\begin{figure}[p]
\plotone{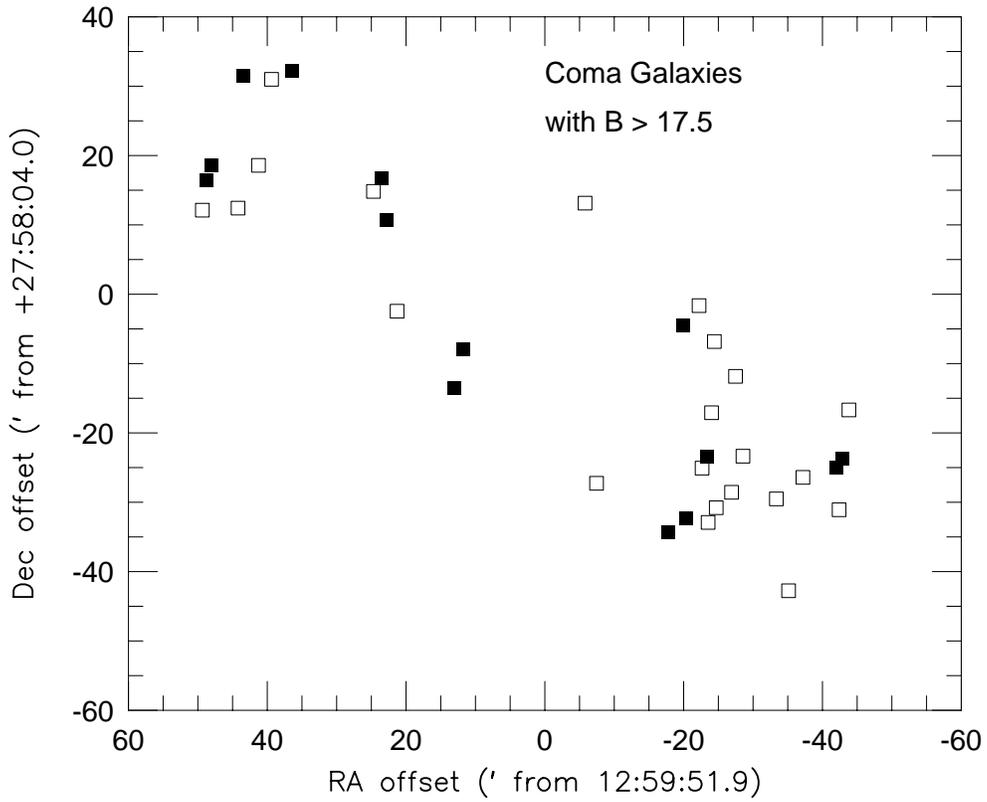}
\figcaption[caldwell.fig9.ps]{Spatial distribution of faint Coma galaxies observed by us
with multi-fiber spectroscopy. The central region has not yet been
well-sampled by our work.  Galaxies with measured 
break amplitude indices are plotted
as filled squares depending if the break amplitude is weak (indicating a
younger population), or as open squares if the break amplitude is normal.
\label{coma_lowl.xy}}
\end{figure}
\clearpage
\begin{deluxetable}{llllrrrrrr}
\tablenum{1}
\tablecolumns{10}
\tablewidth{0pc}
\tablecaption{Galaxy Data}
\tablehead{
\colhead{Galaxy} & \colhead{RA (2000)} &
\colhead{Dec (2000)} & \colhead{B\tablenotemark{a}} &
\colhead{B-R\tablenotemark{a}}  &  \colhead{D\tablenotemark{b}} & \colhead{C93\tablenotemark{c}} &
\colhead{V(CD)\tablenotemark{d}} &\colhead {V(C93)\tablenotemark{e}}
&\colhead{exp.}}
\startdata
GMP2385& 13:00:54.6&   27:50:31.4& 17.62& 1.82&        &S0   &7092 &&3600s\nl
GMP2603& 13:00:33.2&   27:49:27.1& 17.36& 1.80& 83 S0  &S0   &8181&&7200s\nl
GMP3126& 12:59:50.8&   27:49:58.8& 17.55& 1.82&        &S0   &7905&&3600s\nl
GMP3205& 12:59:44.1&   27:52:03.3& 17.61& 1.83&        &S0   &6196&&3600s\nl
GMP3707& 12:59:09.3&   28:02:27.3& 17.76& 1.82&        &S0   &7220&&3600s\nl
GMP4420& 12:58:11.3&   27:56:23.9& 17.60& 1.86&        &E/S0 &8500 &8509&7200s\nl
GMP5284& 12:56:42.2&   27:32:53.6& 17.98& 1.75&        &S0   &7573 &7545&3600s\nl
GMP5320& 13:01:36.4&   27:42:28.7& 18.80& 1.81&        &S0   &7596 &7625&3600s\nl
GMP5362& 12:56:34.0&   27:41:14.4& 17.77& 1.89& 60 E   &S0   &     &6843&10800s\nl
IC3065\tablenotemark{f}&12:12:39.6&   14:42:48  & 14.3 &     &        &S0   &1072 &&1200s\nl
\enddata
\tablenotetext{a}{Data from Godwin et al. (1983)}
\tablenotetext{b}{Numbers and types from Dressler (1980)}
\tablenotetext{c}{Types from Caldwell et al. (1993) and work done therefore}
\tablenotetext{d}{From Colless \& Dunn (1995)}
\tablenotetext{e}{From Caldwell et al. (1993)}
\tablenotetext{f}{Data from Binggeli et al. (1985)}
\label{tab1}
\end{deluxetable}

\begin{deluxetable}{lrrrrrrrr}
\tablenum{2}
\tablecolumns{9}
\tablewidth{0pc}
\tablecaption{Spectral Indices and Errors}
\tablehead{
\colhead{Galaxy} & \colhead{$\rm H\gamma /\lambda 4325$} & 
\colhead{$\rm H\delta /\lambda 4045$} &  \colhead{$\rm H8/3859$} &
\colhead{Hn/Fe I} &  \colhead{Ca II} & \colhead{$\rm H\gamma_{HR}$} &
\colhead{$\rm 4045_{HR}$} & \colhead{$\rm Ca \ I_{HR}$} }
\startdata

GMP2385  &  0.99 &  0.95 &  1.06 &  1.00 &  1.18 &  1.00 &  0.86 &  0.99 \\
  &  $\pm$0.05 &  $\pm$0.05 &  $\pm$0.05 &  $\pm$0.03 &  $\pm$0.00 &  $\pm$0.10 &  $\pm$0.10 &  $\pm$0.10 \\
GMP2603  &  0.82 &  0.81 &  0.78 &  0.80 &  1.18 &  1.29 &  0.82 &  0.96 \\
  &  0.04 &  0.04 &  0.04 &  0.02 &  0.00 &  0.07 &  0.07 &  0.07 \\
GMP3126  &  0.90 &  0.91 &  0.89 &  0.90 &  1.64 &  1.12 &  0.78 &  1.29 \\
  &  0.06 &  0.06 &  0.06 &  0.04 &  0.00 &  0.12 &  0.12 &  0.12 \\
GMP3205  &  0.82 &  0.93 &  0.91 &  0.89 &  1.14 &  1.06 &  0.60 &  0.75 \\
  &  0.07 &  0.07 &  0.07 &  0.04 &  0.00 &  0.14 &  0.14 &  0.14 \\
GMP3707  &  1.06 &  1.04 &  0.91 &  1.00 &  2.03 &  0.79 &  1.00 &  1.15 \\
  &  0.07 &  0.07 &  0.07 &  0.04 &  0.00 &  0.14 &  0.14 &  0.14 \\
GMP4420  &  0.99 &  1.03 &  1.07 &  1.03 &  1.34 &  0.92 &  0.81 &  1.10 \\
  &  0.04 &  0.04 &  0.04 &  0.02 &  0.00 &  0.07 &  0.07 &  0.07 \\
GMP5284  &  0.82 &  0.85 &  0.76 &  0.81 &  0.97 &  1.18 &  0.52 &  0.86 \\
  &  0.05 &  0.05 &  0.05 &  0.03 &  0.00 &  0.10 &  0.10 &  0.10 \\
GMP5320  &  0.92 &  0.88 &  0.77 &  0.86 &  1.34 &  1.13 &  0.71 &  1.16 \\
  &  0.05 &  0.05 &  0.05 &  0.03 &  0.00 &  0.10 &  0.10 &  0.10 \\
GMP5362  &  1.00 &  0.98 &  0.99 &  0.99 &  1.23 &  0.98 &  0.91 &  1.17 \\
  &  0.04 &  0.04 &  0.04 &  0.02 &  0.00 &  0.08 &  0.08 &  0.08 \\
IC3065  &  0.88 &  0.81 &  0.87 &  0.85 &  0.99 &  1.33 &  0.70 &  1.13 \\
  &  0.03 &  0.03 &  0.03 &  0.02 &  0.00 &  0.06 &  0.06 &  0.06 \\
\enddata
\label{tab2}
\end{deluxetable}

\begin{deluxetable}{lrr}
\tablenum{3}
\tablecolumns{3}
\tablewidth{0pc}
\tablecaption{Galaxy Ages}
\tablehead{
\colhead{Galaxy} & \colhead{Hn/Fe I} & \colhead{Age(Gyr)} }
\startdata

GMP2385 & 1.000 & 11.2 \\
GMP2603 & 0.805 & 2.0 \\
GMP3126 & 0.902 & 6.6 \\
GMP3205 & 0.885 & 5.8 \\
GMP3707 & 1.004 & 11.4 \\
GMP4420 & 1.028 & 12.5 \\
GMP5284 & 0.810 & 2.2 \\
GMP5320 & 0.857 & 4.4 \\
GMP5362 & 0.988 & 10.6 \\
VCC140  & 0.854 & 5.8 \\
\enddata
\label{tab3}
\end{deluxetable}

\begin{deluxetable}{lrrr}
\tablenum{4}
\tablecolumns{3}
\tablewidth{0pc}
\tablecaption{Galaxies With Weak Ca II Breaks}
\tablehead{
\colhead{Galaxy} & \colhead{RA} & \colhead{Dec} & \colhead{V$_\odot$}  \\
\colhead{} & \multicolumn{2}{c}{(2000)} & \colhead{(\kms)} }
\startdata
 & \multicolumn{2}{c}{SW Region} & \\
GMP4215 & 12:58:31.5 & 27:23:41.6 & 7554 \\
GMP4314 & 12:58:21.8 & 27:53:32.2 & 7252 \\
GMP4330 & 12:58:20.4 & 27:25:46.1 & 7601 \\
GMP4469 & 12:58:06.7 & 27:34:37.1 & 7452 \\
GMP5284 & 12:56:42.2 & 27:32:53.6 & 7545 \\
GMP5320 & 12:56:38.3 & 27:34:15.4 & 7625 \\
 & & & \\
 & \multicolumn{2}{c}{Central Region} & \\
GMP2421 & 13:00:50.9 & 27:44:34.5 & 8126 \\
GMP2478 & 13:00:45.3 & 27:50:08.0 & 8720 \\
 & & & \\
 & \multicolumn{2}{c}{NE Region} & \\
GMP864  & 13:01:11.8 & 28:26:06   & 6573 \\
GMP885  & 13:01:09.5 & 28:30:27   & 8147 \\
GMP1107 & 13:00:45.7 & 28:45:28   & 5412 \\
GMP1379 & 13:00:13.9 & 28:46:13   & 5756 \\
GMP1975 & 12:59:10.7 & 28:24:56   & 5272 \\

\enddata
\label{tab4}
\end{deluxetable}

\begin{deluxetable}{lrrr}
\tablenum{5}
\tablecolumns{3}
\tablewidth{0pc}
\tablecaption{Galaxies With Normal Ca II Breaks}
\tablehead{
\colhead{Galaxy} & \colhead{RA} & \colhead{Dec} & \colhead{V$_\odot$}  \\
\colhead{} & \multicolumn{2}{c}{(2000)} & \colhead{(\kms)} }
\startdata
 & \multicolumn{2}{c}{SW Region} & \\
GMP3588 & 12:59:18.3 & 27:30:48.4 & 6076 \\
GMP4420 & 12:58:11.3 & 27:56:23.9 & 8509 \\
GMP4447 & 12:58:09.7 & 27:32:57.7 & 6958 \\
GMP4479 & 12:58:05.9 & 27:25:07.7 & 5749 \\
GMP4502 & 12:58:03.4 & 27:40:56.6 & 7209 \\
GMP4519 & 12:58:01.4 & 27:51:12.6 & 5638 \\
GMP4535 & 12:58:00.6 & 27:27:14.4 & 7653 \\
GMP4630 & 12:57:50.6 & 27:29:27.2 & 7335 \\
GMP4656 & 12:57:47.7 & 27:46:10.0 & 5764 \\
GMP4714 & 12:57:43.1 & 27:34:39.3 & 7226 \\
GMP4956 & 12:57:21.5 & 27:28:28.8 & 6902 \\
GMP5012 & 12:57:13.9 & 27:15:13.4 & 5174 \\
GMP5102 & 12:57:04.2 & 27:31:33.2 & 8328 \\
GMP5296 & 12:56:40.8 & 27:26:51.7 & 7310 \\
GMP5362 & 12:56:34.0 & 27:41:14.4 & 6843 \\
 & & & \\
 & \multicolumn{2}{c}{Central Region} & \\
GMP3481 & 12:59:25.6 & 28:11:12.9 & 7718 \\
 & & & \\
 & \multicolumn{2}{c}{NE Region} & \\
GMP1076 & 13:00:48.4 & 28:26:27   & 7243 \\
GMP1201 & 13:00:35.2 & 28:32:37   & 7142 \\
GMP1266 & 13:00:27.1 & 28:45:03   & 7007 \\
GMP1885 & 12:59:19.7 & 28:28:59   & 7808 \\
GMP2033 & 12:59:03.8 & 28:11:43   & 7491 \\
\enddata
\label{tab5}
\end{deluxetable}

%

\end{document}